\documentclass[twocolumn,pre,superscriptaddress]{revtex4-2}
\usepackage[utf8]{inputenc}
\usepackage{amsmath}
\usepackage{amssymb}
\usepackage{physics}
\usepackage{graphicx}
\usepackage{multirow}
\usepackage{placeins}

\usepackage{xcolor}
\usepackage{bm}
\usepackage{xcolor,cancel}
\usepackage{lipsum}
\usepackage{soul}
\setcitestyle{super}

\newcommand{\revise}[1]{{\color{black} #1}} 
\newcommand{\revfinal}[1]{{\color{black} #1}} 

\newcommand{\mc}[1]{{\color{black} #1}} 
\newcommand{\fer}[1]{{\color{black} #1}} 
\newcommand{\pg}[1]{{\color{black} #1}} 
\newcommand{\mcm}[1]{{\color{black} #1}}

\newcommand{\pgf}[1]{{\color{black} #1}} 

\newcommand{\Qbf}{\mathbf{Q}}

\newcommand{\rbf}{\mathbf{r}}

\newcommand{\vbf}{\mathbf{v}}

\newcommand{\sbf}{\boldsymbol{\sigma}}

\newcommand{\eq}{\begin{equation}}
\newcommand{\eqq}{\end{equation}}

\begin{document}
\title{\revise{Traveling waves at the surface of active liquid crystals}}
\author{Paarth Gulati}
\affiliation{Department of Physics, University of California Santa Barbara, Santa Barbara, CA 93106, USA}
\author{Fernando Caballero}
\affiliation{Department of Physics, University of California Santa Barbara, Santa Barbara, CA 93106, USA}
\affiliation{Department of Physics, Brandeis University, Waltham, Massachusetts 02453, USA}
\author{Itamar Kolvin}
\affiliation{School of Physics, Georgia Institute of Technology, Atlanta, GA 30332, USA}
\author{Zhihong You}
\affiliation{Fujian Provincial Key Laboratory for Soft Functional Materials Research, Research Institute for Biomimetics and Soft Matter, Department of Physics, Xiamen University, Xiamen, Fujian 361005, China}
\author{\mc{M.} Cristina Marchetti}
\affiliation{Department of Physics, University of California Santa Barbara, Santa Barbara, CA 93106, USA}

\date{\today}

\begin{abstract}
Active liquid crystals exert nonequilibrium stresses on their surroundings through constant consumption of energy, giving rise to dynamical steady states not present in equilibrium. 
The paradigmatic example of an active liquid crystal is a suspension of microtubule bundles powered by kinesin motor proteins, which exhibits self-sustained spatiotemporal chaotic flows. This system has been  modelled using continuum theories that couple the microtubule orientation to active flows. 
Recently the focus has shifted to the interfacial properties of mixtures of active liquid crystals and passive fluids. Active/passive interfaces have been shown to support propagating capillary waves in the absence of inertia and offer a promising route for relating experimental parameters to those of the continuum theory. 
In this paper we report the derivation of a minimal model that captures the linear dynamics of the interface between an active liquid crystal and a passive fluid. 
We show that the dynamics of the interface, although powered by active flows throughout the bulk, is qualitatively captured by equations that couple non-reciprocally interface height and nematic director at the interface. This minimal model reproduces the dynamical structure factor evaluated from numerical simulations and the qualitative form of the wave dispersion relation seen in experiments. 
\end{abstract}

\maketitle

\section{Introduction}

Active liquid crystals have become a paradigmatic example of active matter, with nematic order observed on many scales in both living and 
engineered systems. Examples include epithelial tissue and suspensions of reconstituted proteins extracted from living cells, such as microtubule-kinesin \cite{sanchez2012spontaneous,sanchez2011cilia,lemma2021multiscale,needleman2017active} and actomyosin fluids \cite{lee2021active,murrell2015forcing,wang2012active}. The active units in these systems are elongated  apolar complexes powered by the hydrolysis of adenosyn-triphospate (ATP) to exert forces on their environment and drive self-sustained, spatiotemporally chaotic flows. 

Recently, the realization of immiscible mixtures of active liquid crystals and passive fluids has allowed the study of the effect of activity on  liquid-liquid phase separation (LLPS)~\cite{adkins2022dynamics,tayar2023controlling}. Both experiments and numerical work have revealed a wealth of new interfacial phenomena, including giant activity-driven interfacial fluctuations, traveling capillary waves in the absence of inertia, and striking wetting behavior where the active fluid can climb the walls against gravity~ \cite{adkins2022dynamics,caballero2022activity,tayar2023controlling,kempf2019active,bhattacharyya2024phase,bhattacharyya2023phase,blow2014biphasic}.

Continuum theories have proven to be an effective framework for describing the long wavelength behavior of active liquid crystals~\cite{marchetti2013hydrodynamics,simha2002hydrodynamic,doostmohammadi2018active}. In these theories activity enters through active forces driven by deformation of \mc{the orientation of the elongated active units,} which in turn drives active flows. The feedback between deformations and flow generically destabilizes the homogeneous \mc{ordered} state ~\cite{simha2002hydrodynamic}, generating \mc{self-sustained} flows. Recent extensions to mixtures of active and passive fluids have shown that the coupling of flow and liquid crystalline degrees of freedom additionally modifies the phase separation and its kinetics, and can also capture many of the observed interfacial phenomena~\cite{adkins2022dynamics,caballero2022activity,tayar2021active,bhattacharyya2024phase,bhattacharyya2023phase,blow2014biphasic,kempf2019active,doostmohammadi2018active}.

Here we present an analytical derivation of the linear dynamics of  interfacial fluctuations in a phase separated mixture of an active liquid crystal and a passive fluid. 
We show that the observed capillary waves can be described by a minimal model that couples fluctuations in the interface height to director fluctuations at the interface. We examine both the situations where, in the absence of activity, the liquid crystal is in the nematic state and in the isotropic state. Although bulk nematic fluids are generically unstable,  the interface can stabilize the phase separated mixture through traveling surface waves in a range of intermediate wavenumbers. Our minimal model captures qualitatively the experimental observations. We also evaluate numerically the dynamical structure factor of interfacial fluctuations, which clearly displays traveling waves, and show that  
it agrees qualitatively with the analytical model. Active capillary waves arise generically via a non-reciprocal coupling of the interface height and the liquid crystal director field, mediated by hydrodynamic flows through the bulk. \pgf{With a nematic base state, t}he dispersion relation has the generic form recently \mc{predicted} in a new class of non-reciprocal phase transitions observed in coupled conserved fields~\cite{you2020nonreciprocity,saha2020scalar,frohoff2023nonreciprocal,frohoff2023non,brauns2023non} or in a conserved field coupled to a Goldstone mode~\cite{fruchart2021non}. It arises when the eigenvalues of the matrix describing the dynamical evolution of the system coalesce at an exceptional point at the edge of a band of linearly unstable modes~\cite{adkins2022dynamics}. 

Previous analytical work~\cite{adkins2022dynamics} has focused on the calculation of the spectrum of equal-time interfacial fluctuations which can be well approximated by effectively neglecting the dynamics of the director angles. In contrast, we show here that it is essential to incorporate the director dynamics in order to understand the origin of capillary waves. 
Soni \emph{et al.}~\cite{soni2019stability} have also explicitly obtained the dispersion relation of capillary waves for the case of isotropic liquid crystals, but their method tends to  obscure the physical mechanism that drives the travelling waves.

The effect of activity on phase separation has been studied extensively with scalar models of a phase separating concentration field advected by flow~\cite{tiribocchi2015active,cates2018theories,cates2019active}. In this case activity is introduced via an active capillary stress of strength free to differ from the equilibrium interfacial stiffness appearing in the free energy, allowing for the possibility of an effective negative interfacial tension, which can qualitatively modify the coarsening dynamics and the phase separation.
On the other hand, the \pgf{active capillary} stress is proportional to gradients of the concentration field and only acts on the
interfaces of the phase separated regions. Here, in contrast, we find that that the emergence of active capillary waves requires the feedback from active stress throughout the bulk fluid, as induced by the coupling of flow to liquid crystalline degrees of freedom.

The remainder of the paper is organized as follows. In Section ~\ref{sec:continuum_model} we introduce the continuum model of the active/passive fluid mixture. In Section~\ref{sec:linear_theory}, we derive linear equations for height and director fluctuations around an initially flat interface with projections methods used before in the literature \cite{bray2001interface, besse2023interface, caballero2022activity}. We carry out the calculation of the linear interfacial dynamics in two regimes, corresponding to situations where the liquid crystal is in the isotropic or nematic regime when passive. We show that the linear model captures the qualitative feature of the dispersion relation of active capillary waves in both regimes.   We also discuss the role of gravity in suppressing long wavelength instabilities while preserving propagating modes \mc{in Section} \ref{sec:gravity}.
Finally, in Sec.~\ref{sec:numerics}, we solve numerically the full continuum equations to compute the dynamical structure factor of the interface, which clearly displays propagating modes at intermediate wavenumbers, in qualitative agreement with the linear theory. 
We conclude with a few remarks and open questions.

\section{Continuum Model}
\label{sec:continuum_model}
We consider a phase separating \mcm{mixture} of an isotropic passive Newtonian fluid and an active \mcm{liquid crystal} in two dimensions, thus with one-dimensional interfaces.

\mcm{We use a continuum model with} a phase field $\phi$ describing the local \mcm{relative concentration} of
active 
and passive 
fluid \mcm{and a velocity field $\mathbf{v}$ capturing fluid flow.} \mcm{Liquid crystalline order is described by the} nematic \mcm{alignment tensor}, $Q_{ij} = S (n_in_j -\delta_{ij}/2)$ where $S$ is the \mcm{order parameter} and $\mathbf{n}$ is the local director field \mcm{- a unit vector pointing along the direction of broken symmetry}.
The dynamics is governed by the following equations,
\begin{align}
    D_t \phi &= M \nabla^2\mu\;,\label{eq:phi_full}\\
    D_t \mathbf{Q} &= \lambda \mathbf{A} - [\boldsymbol{\Omega},\mathbf{Q}] + \frac{1}{\gamma}\mathbf{H}\;,\label{eq:Q_full}\\
    0 &= \eta \nabla^2\mathbf{v} +  \mcm{\boldsymbol\nabla \cdot \left(\boldsymbol{\sigma}^e +\boldsymbol{\sigma}^\phi+\boldsymbol{\sigma}^a\right)}  - \mcm{\boldsymbol\nabla} P\;, \label{eq:StokesFlow}
\end{align}
where $D_t = \partial_t + \mathbf{v}\cdot\nabla$ is the material derivative. 

The phase field $\phi$ obeys relaxational dynamics with \mcm{mobility $M$, \mc{assumed constant,}} and chemical potential \mcm{$\mu = \delta F_\phi/ \delta \phi$}
determined by a \pgf{Ginzburg-Landau} free energy
\begin{align}
    F_\phi = \int d\mathbf{r}\,\left[ f_0 (\phi) + 2\kappa (\nabla\phi)^2\right]\;.
    \label{eq:Fphi}
\end{align}

The free energy density for the uniform state is taken to be of the form, $f_0(\phi) = -4 a \phi^2(1-\phi)^2,$ with $a<0$, which corresponds to a bistable system with ground states
representing the isotropic passive fluid ($\phi=0$) and the active liquid crystal ($\phi=1$).

The nematic tensor $\Qbf$ relaxes according to a molecular field  $H_{ij} = - \delta F_Q /\delta Q_{ij}$, derived from the Landau-de Gennes free energy
\begin{align}
 F_Q = \int d\mathbf{r}\,\left( \frac{r \phi}{2} \Tr(\mathbf{Q}^2)  + \frac{u}{4} (\Tr\mathbf{Q}^2)^2 + \frac{K}{2} (\partial_k Q_{ij})^2\right)\;,
    \label{eq:FQ}
\end{align}
with $\gamma$ the rotational viscosity.  It additionally couples to the flow through the rate of strain tensor $A_{ij} = (\partial_iv_j + \partial_jv_i\mcm{-\delta_{ij}\nabla\cdot\mathbf{v}})/2$  and vorticity tensor $\Omega_{ij} = (\partial_iv_j - \partial_jv_i)/2$.

\mcm{The flow velocity $\vbf$ is determined by Eq.~\eqref{eq:StokesFlow} that balances
 viscous dissipation with viscosity $\eta$, gradients of pressure $P$ and elastic, capillary and active stresses.}
 We further assume incompressible flow ($\boldsymbol\nabla\cdot\vbf =0$) in the Stokes limit.
The \mcm{capillary force, $\boldsymbol\nabla\cdot\sbf^\phi=-\phi\boldsymbol\nabla\mu$, arises from} the free energy cost of creating interfaces.
One can show that the capillary stress can be written as ~\cite{cates2018theories}
\begin{align}
    \sigma^{\phi}_{ij} = -\kappa\left[\revise{\left(\partial_i\phi\right)\left(\partial_j\phi\right)} - \frac{1}{2}\delta_{ij}(\nabla\phi)^2\right]\;.
    \label{eq:sigmaphi}
\end{align} 
The 
liquid-crystalline degrees of freedom 
\mcm{are additionally responsible for elastic stress $\sbf^e$ and active stress $\sbf^a$, given by}\pgf{
\begin{equation}
\begin{aligned}
\sigma^e_{ij} &= -\lambda H_{ij}+[\mathbf{Q}, \mathbf{H}]^{\rm{A}}_{ij} \;,\\
\sigma^a_{ij} &= \alpha \phi Q_{ij}\;,
\label{eq:sigmaQ}
\end{aligned}
\end{equation}
where $[\mathbf{Q}, \mathbf{H}]^{\rm{A}}_{ij}=Q_{ik}H_{kj}-H_{ik}Q_{kj}$ denotes the antisymmetrized product of any two tensors. The parameter $\alpha,$ with dimensions of stress, controls the activity and depends on the  biomolecular processes that drive the flows. Its sign is determined by whether active forces are extensile ($\alpha<0$), as considered here, or contractile ($\alpha>0$). 
}

\section{Linear interfacial dynamics}
\label{sec:linear_theory}

\begin{figure}
    \centering
    \includegraphics[width=0.4\textwidth]{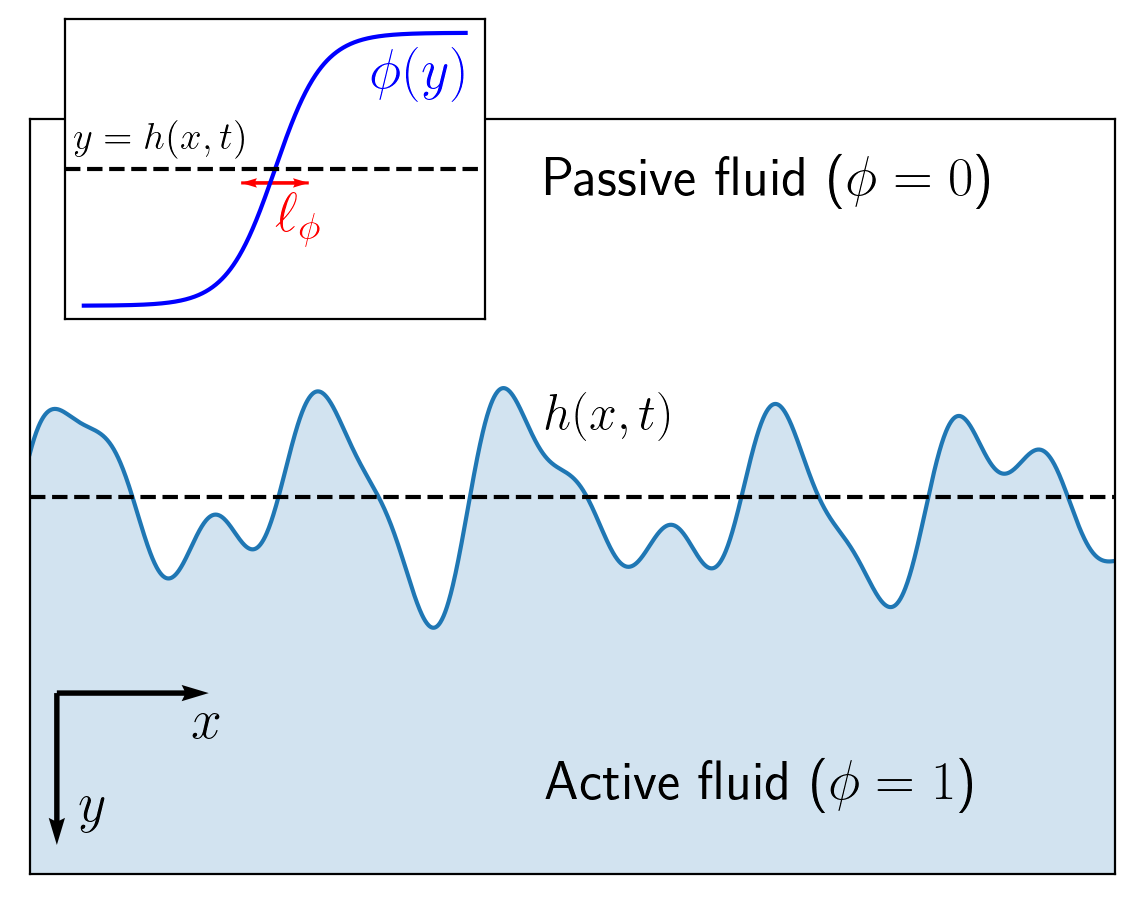}
    \caption{Sketch of the phase separated active/passive fluid. The blue line is the interface between the active and passive phases. \revfinal{The height fluctuations have been exaggerated for illustrative purposes.} Inset:  profile of the $\phi$ field across the interface showing the interfacial thickness $\ell_\phi$.} 
    \label{fig:schematic}
\end{figure}

In the following we examine the stability and dynamics of the interface separating the bulk phases of the two fluids \revise{in an infinite plane}. We assume the active liquid crystal ($\phi=1$) occupies the region $y>0$ and the passive fluid ($\phi=0$) occupies the region $y<0$. 
The profile of the flat interface separating the two regions at $y=0$ is easily found from the solution of \revise{$\delta F_\phi/\delta\phi= 0$, as sketched in Fig. \ref{fig:schematic}. The solution describes a smooth function varying between the bulk values of $\phi$ across a length scale $\ell_\phi=\sqrt{\kappa/|a|}$. We will consider a sharp interface approximation, assuming $\ell_\phi$ is much smaller than any other length scale.}
\fer{We stress that we have made the nonstandard choice of the $y$ axes pointing down in Fig. \ref{fig:schematic}. This choice is made purely to obtain algebraically cleaner expressions for the continuum equations for the interface, and at the same time to display the active flow below the passive one (as is the case in experimental realizations of this form of phase separation, where the active phase is heavier).}

Below we adapt the method first described in Ref. \cite{bray2001interface} to derive linear equations for the dynamics of activity-driven fluctuations in the interface height $h(x,t)$ \mc{and director angle $\theta(\mathbf{r},t)$ defined by $\mathbf{n}=(\cos\theta,\sin\theta)$}(see additional details in Appendix \ref{app_sec:interfaceprojection}).
 
We  write 
\begin{align}
    \phi(x,y,t) = g({y - h(x,t)})\;,
    \label{eq:h}
\end{align} 
where for an infinitely  sharp interface $g(y)$ is well approximated by a step function, i.e., $g(y)=\Theta(y)$. 
\mcm{We consider the regime where the dynamics of the phase field is dominated by advection by flow} and ignore the right hand side of \mcm{Eq.~\eqref{eq:phi_full}}. \mcm{Inserting the ansatz given in Eq.~\eqref{eq:h}, we can then write} 
\eq
    { \partial_t\phi= - g'(y -h(x,t) )\partial_t h(x,t) = -v_i \partial_i g(y -h(x,t))}\;,
\eqq
where the prime denotes a derivative $g'(y)=dg/dy$. This can be simplified by integrating across the interface and using the sharp interface approximation ($g'(y) \approx \delta(y))$, to give to linear order
\eq
\partial_t h (x, t) = v_y(x,y=0,t)
\;.\label{eq:h_response}
\eqq

Next, we eliminate the flow and project the dynamics onto the interface to obtain equations  for height fluctuations and fluctuations of the liquid crystalline degrees of freedom at the interface
~\cite{bray2001interface,besse2023interface,caballero2022activity}.
This can be done by formally solving Stokes equation, i.e. writing the flow as
\begin{align}
    v_i(\mathbf{r}) = \int d\mathbf{r'}\, T_{ij}(\mathbf{r -r'})&f_j(\mathbf{r'}) \;, \label{eq:flow_generic} 
\end{align}
where $T_{ij}(\mathbf{r})$ are the components of the Oseen tensor, which in Fourier space and for a bulk system are given by
\eq
    \tilde{T}_{ij}(\mathbf{k}) = \frac{1}{\eta |\mathbf{k}|^2}\left(\delta_{ij}-\frac{k_ik_j}{k^2}\right)\;,
    \label{eq:Tq}
\eqq
and where the driving force for the flow $f_i(\mathbf{r})$ 
contains both equilibrium and active stresses,
\begin{align}
   f_i(\mathbf{r})
=\partial_j \left(\sigma^\phi_{ij}+\sigma^a_{ij}\right),
\label{eq:fi}
\end{align}
\pgf{where, for simplicity, we have ignored the elastic stress $\boldsymbol{\sigma}^e$ which is higher order in gradients of the nematic field.}

\mcm{It is convenient to write} 
Eq.~\eqref{eq:flow_generic} as 
$\vbf = \vbf^\phi+\vbf^a$. \mcm{The explicit form of these two contributions is then} 
\begin{align}
    v_i^\phi(\mathbf{r},t) &= \int d\rbf' T_{ij}(\rbf - \rbf') \mu(\mathbf{r}')\mcm{\partial'_j\phi(\mathbf{r}')} \;,\\
     v_i^\mcm{a}(\mathbf{r}, t) &= \int d\mathbf{r'} T_{ij}(\mathbf{r-r'}) \left[ \alpha \partial'_k\left( \phi(\rbf ') Q_{jk}(\mathbf{r'}, t)\right) \right] \;.\label{eq:vQ}
\end{align}
Here, we have written the capillary force coming from the stress in Eq. \ref{eq:sigmaphi} as $f_i^\phi = \mu\revise{\partial}_i \phi$, which can be done by an integration by parts due to incompressibility. \mcm{In the remainder of this section we evaluate the flows induced by height and director fluctuations to obtain a closed description of the interfacial dynamics. }

\subsection{\mcm{Flows driven by passive capillary stresses}}
 Following previous work \cite{bray2001interface}, the flow $v_i^\phi$ can be calculated in terms of the perturbation $h(x,t)$ by using Eq.~\eqref{eq:h} and expanding for small $h$ as $\partial_i\phi\simeq -g'(y)\partial_xh$. The expre\revise{s}sion for the flow velocity can then be evaluated taking advantage of the sharpness of the function $g'$, with the result
\eq\label{eq:vphi}
v_i^\phi(\mathbf{r},t) = {\sigma} \int dx' T_{iy}(x-x', y) \partial_{{x'}}^2 h(x', t)\;, 
\eqq
where we have integrated along the coordinate $y$ perpendicular to the interface (see algebraic details in Appendix \ref{app_sec:interfaceprojection}). Here $\sigma$ is the interface tension, defined as the energy cost per unit length of creating density gradients. It can \revise{easily be calculated analytically using the equilibrium solution for the interface profile} and is given by
\begin{align}
    \sigma = 
    2\kappa \displaystyle\int_{-\infty}^\infty dy'\left[\nabla'\phi(y')\right]^2 = \sqrt{8\kappa |a|/9}\;.
    \label{eq:sigma}
\end{align}

It is convenient to Fourier transform in $x$ and define the partial transform of the Oseen tensor as 
\begin{align}
    G_{ij}(k,y) = \int_{-\infty}^{\infty}\, dk_y e^{i k_y y} T_{ij}(k, k_y)\;.
\end{align}
The passive flow is then given by
\begin{align}
    v_i^\phi(k, y) &=  {-\sigma} k^2 G_{iy}(k,y) h(k, t)\;. \label{eq:v_interface_h}
\end{align}
or explicitly,
\begin{align}
    v_x^\phi(k, y)&= \frac{\sigma|k|}{4\eta}e^{-|ky|}iky~ h(k, t) \;,\\
    v_y^\phi(k, y)&= -\frac{\sigma |k|}{4\eta}(1+|ky|)e^{-|ky|}  h(k, t)\;.
\end{align}
Finally, substituting in Eq.~\eqref{eq:h_response}, we obtain
\begin{align}
    \partial_t h(k,t)&=-\frac{\sigma |k|}{4\eta} h(k,t)+v_y^a(k,y=0)\;.
    \label{eq:hphi}
\end{align}
In the absence of active flows ($\mathbf{v}^a =0$), this describes the hydrodynamic relaxation of interface fluctuations due to capillary stresses.

\subsection{Flows driven by active stresses}
To evaluate the contribution to velocity fluctuations from active stresses \mc{and} close Eq.~\eqref{eq:hphi}, we need to examine the linear dynamics of fluctuations of the nematic tensor at the interface. To do this we will consider separately the case where the liquid crystal is in a nematic or isotropic state in the absence of activity.

The case of an isotropic passive liquid crystal is directly relevant to recent experimental realizations of active/passive phase separating mixtures, 
where the concentration of microtubules in the active phase is too low for the onset of nematic order to occur in absence of activity. On the other hand, both simulations and experiments have shown that in this regime
local nematic order is created by active stresses \mcm{through flow alignment}~\cite{adkins2022dynamics,tayar2021active,lemma2022active,srivastava2016, santhosh2020activity}.

\subsubsection{\textbf{Ordered Nematic}}
\label{subsec:nematic}

In this section we consider the case in which the liquid crystal is in the nematic state in the absence of activity. \mcm{We note that bulk extensile active} nematics are \mcm{generically} unstable to \mcm{bend fluctuations at}  any nonzero activity~\cite{simha2002hydrodynamic}.
\mcm{It is nonetheless instructive to examine the interfacial dynamics and ask whether the interface has any stabilizing effect.}

We assume that in the bulk phase separated state  the director is aligned along the direction of the flat interface separating active and passive fluid at $y=0$ ($\mathbf{n}_0=\mathbf{\hat{x}}$). \revise{This assumption is justified by the fact that
extensile active stresses are known to create an effective alignment interaction that aligns the director with the interface \cite{blow2014biphasic, coelho2023active, zhao2024asymmetric}. In a contractile active system, the shear flows tend to align the director perpendicular to the interface. The dispersion relations for such a system can be analogously derived by assuming normal anchoring for the base state}.
In addition, for length scales large compared to both the interfacial thickness and the nematic correlation length, we simply slave the magnitude of the order parameter to the field $\phi$, i.e., $S_0(y)\simeq\phi(y)$. 
The base state is then described by $Q_{xx}^0=-Q_{yy}^0=S_0(y)$, $Q_{xy}^0=0$ and zero flow velocity.
To examine the dynamics of small fluctuations, we then retain only terms linear in the  interfacial height  $h(x,t)$ and the director angle $\theta(x,y,t)$ and neglect fluctuations in the magnitude of the order parameter, as appropriate deep in the nematic state. This gives $\delta Q_{xx}=-\delta Q_{yy}\simeq -g'(y)h(x,t)$ and $\delta Q_{xy}\simeq g(y) \theta(x,y,t)$. Inserting these expressions in Eq.~\eqref{eq:activeflows_general} and taking a Fourier transform in $x$ we obtain
\begin{align}
    &v_i^a(k,y) = -\alpha h(k,t) \left[ik G_{ix}(k, y) \mcm{-} \partial_y G_{iy}(k,y)\right] S_0(0)\notag \\ &+ \alpha\int_0^{\infty} dy'\;\left[ ik G_{iy}(k, y-y')-\partial_{y'} G_{ix}(k, y-y') \right]  \theta(k,y', t)\;.
    \label{eq:vi_act}
\end{align}

The linear dynamics of director angle fluctuations is obtained by linearizing Eq.~\eqref{eq:Q_full} around the base state and is given by (for $y\geq0$)
\begin{align}
            \partial_t \theta(k,y,t) = \frac{\lambda+1}{2}ikv_y +\frac{\lambda-1}{2}\partial_yv_x
            - D (k^2-\partial_y^2)\theta\;,\label{eq:theta_lin}
\end{align}
where $D= K/\gamma$.

Our goal is to obtain a minimal model that describes interfacial dynamics in terms of closed equations for height fluctuations $h(x,t)$ and director fluctuations at the interface, $\theta_0(x,t)=\theta(x,y=0,t)$. To achieve this we need to make some assumption on the dependence of angle fluctuations on the distance $y$ from the interface. It is clear from Eqs.~\eqref{eq:vi_act} that fluctuations throughout the bulk fluid up to distances of order $1/k$ contribute to the right hand side of Eqs.~\eqref{eq:vi_act}, hence become most important at small wavenumbers. On the other hand, our goal is to develop a minimal model that captures the capillary waves observed in experiments in a regime of intermediate wave numbers~\cite{adkins2022dynamics}. 
We then first simply ignore the $y$ dependence and assume $\theta(k,y,t)=\theta_0(k,t)$ in Eq.~\eqref{eq:vi_act}. This assumption yields the following coupled equations for the interfacial dynamics:
\begin{align}
    \partial_t h &= -\frac{\sigma}{4\eta}\abs{k} h + \frac{\alpha}{2\eta k^2} i k\theta\;,\label{eq:hthetaordering_1}\\
    \partial_t \theta &= -\frac{\sigma \abs{k}}{4\eta} ikh - \left(\frac{\alpha\lambda}{2\eta} + D k^2 \right) \theta\;.\label{eq:hthetaordering_2}
\end{align}
\begin{figure}[h]
    \centering
\includegraphics[width=0.49\textwidth]{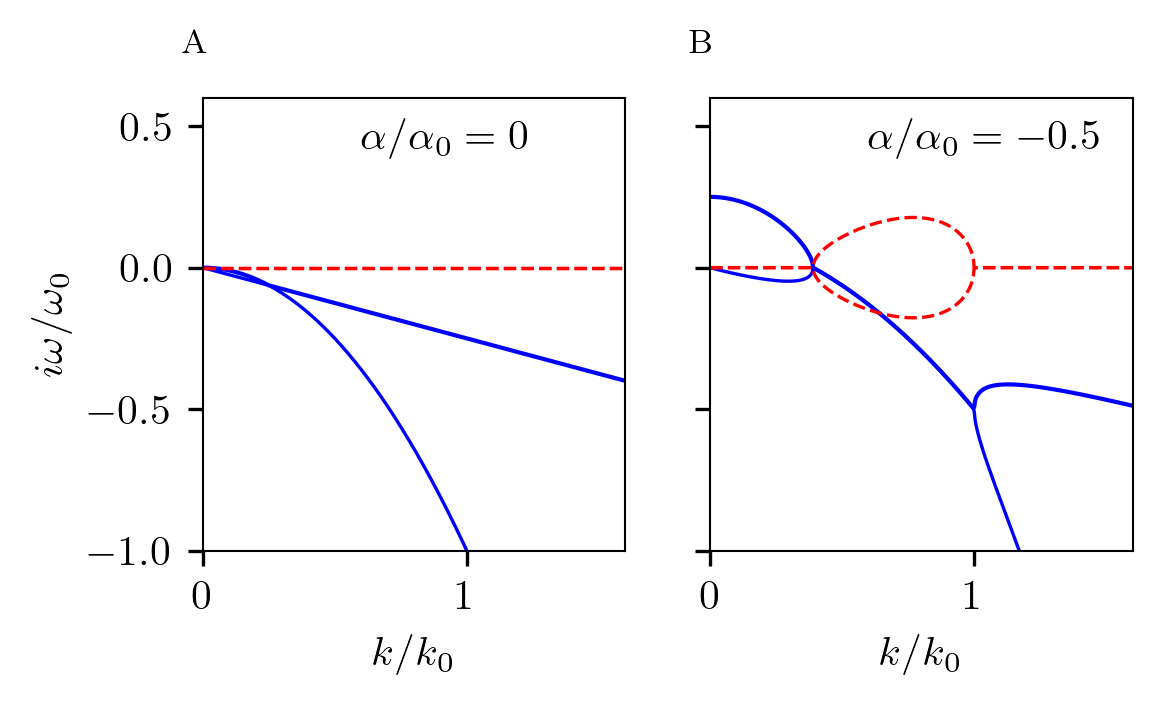}
\caption{Dispersion relation for the modes obtained from Eqs. \eqref{eq:hthetaordering_1} and \eqref{eq:hthetaordering_2}, with $\lambda=1.0$. The blue solid lines show the real part of the eigenvalues and the red dashed lines are the imaginary parts. With our convention, fluctuations decay when  $\rm{Re}\{i\omega\} < 0$ and grow when $\rm{Re}\{i\omega\} > 0$. The frequency, wavenumber and activity are written in terms of $\omega_0 = \sigma^2/(\eta^2 D)$, $k_0 = \sigma/(\eta D)$ and $ \alpha_0 = \sigma^2/(\eta D)$, which arise as the natural units from Eqs.~\eqref{eq:hthetaordering_1} and \eqref{eq:hthetaordering_2}. (A) For $\alpha =0$, the two modes are decoupled (and intersect at $k =k_0/4$). (B) For extensile activity, $\alpha<0$, the modes coalesce and develop a band of propagating modes. At small $k$ we see the instability of the active nematic reflected in the positive real part of $i\omega$.}
\label{fig:nematic_disp}
\end{figure}
The dispersion relation of the modes describing the dynamics of height and director fluctuations at the interface can be obtained from the eigenvalues of the matrix $M$ defined by Eqs.~\eqref{eq:hthetaordering_1} and \eqref{eq:hthetaordering_2} such that $\partial_t\binom{h}{\theta} = M \binom{h}{\theta}$. The real and imaginary parts of the dispersion relations are shown in Fig.~\ref{fig:nematic_disp}. 
In absence of activity the equations are decoupled and the modes are always real, describing decay of height fluctuations due to interfacial tension and of director fluctuations due to nematic stiffness (Fig.~\ref{fig:nematic_disp}\;A). A finite value of activity drives the instability of director fluctuations at small wavenumber, produced by the well-known generic bend instability of extensile active nematics~\cite{simha2002hydrodynamic}. 

For finite extensile activity, the modes are coupled \mc{and display an unstable band of wavenumbers associated with the generic instability of the bulk system. at $k\sim k^*=\sigma/4\eta D$ the modes coalesce. and become complex, signaling} 
 the onset of propagating capillary waves 
 \mc{which persist} in a band of wavenumbers, $(k_{\text{min}},k_{\text{max}})$. To lowest order in activity, one can write $(k_{\text{min}},k_{\text{max}}) = (k^*-\delta k, k^* +\delta k)$, with $\delta k = \sqrt{2|\alpha|/\eta D}$. The range of wavenumbers over which one observes traveling waves is controlled by length scale over which a deformation of the director can diffuse in an active time $\eta/|\alpha|$ and grows with activity, consistent with simulations and experiments~\cite{adkins2022dynamics}.

At finite activity the dispersion relation obtained from Eqs.~\eqref{eq:hthetaordering_1} and~\eqref{eq:hthetaordering_2} has the generic form obtained for two nonreciprocally coupled conserved fields or a conserved field coupled to a Goldstone mode \cite{you2020nonreciprocity,saha2020scalar,brauns2023non,fruchart2021non}. Travelling waves emerge near the edge on the band of unstable modes where the eigenvalues and eigenvectors coalesce and the matrix $M$ becomes nondiagonalizable. Here the  nonreciprocal coupling between  bend deformations of the director field and passive restoring forces from interfacial tension is responsible for the onset of traveling waves. \mc{Director fluctuations drive interfacial fluctuations, which in turn tend to suppress orientational deformations. The coupling is mediated by active flows through} a sort of ``delayed anchoring'' of the director field at the interface: height fluctuations drive director deformations that push height fluctuations before they have had the chance to relax or grow.

By slaving the dynamics of the angle $\theta$ to interfacial fluctuations, i.e., assuming $\theta\sim \partial_xh$, one recovers the simple form of the generic bend instability of an active liquid crystal. In this case, the equation for height fluctuations takes the simple form
\begin{align}
\partial_t h &= -\frac{\sigma}{4\eta}\abs{k} h - \frac{\alpha}{2\eta} h\;,
\end{align}
and of course does not display any traveling waves since we have eliminated the dynamics of one of the degrees of freedom. In this limit, height fluctuations are generically unstable at all values of extensile activity ($\alpha<0$) as a result of the generic bend instability of extensile active nematics. In other words, the interface breaks up, as observed in mixtures of active liquid crystals and passive fluids \cite{tayar2023controlling,caballero2022activity}. 

Finally, the  qualitative features of the dispersion relation are quite robust and independent of the form assumed for the $y$ dependence of director angle fluctuations when evaluating the corresponding contribution to active flows. 
We show in Appendix \ref{app_sec:ordered_nematic} that the qualitative structure of the dispersion relation is unchanged if we assume an exponential decay $\theta(k,y,t)=\theta_0(k,t)e^{-|y|/\ell}$, with $\ell$ the correlation length of active flows. This form is motivated by the fact that the active liquid crystal develops chaotic turbulent flows  at any finite activity. A more detailed calculation that couples height fluctuations to moments of angle fluctuations weighted by the full $\sim e^{-|k||y|}$ Green's function also reproduces the same qualitative structure of the dispersion relations.

\subsubsection{\textbf{Isotropic Liquid Crystal}}
Active liquid crystals reconstitued from cell extracts, such as microtubule-kinesin suspensions, are often in the isotropic phase when passive, corresponding to $r>0$, with nematic order only built-up locally by the coherent flows generated by active stresses. In this case the base state for an active/passive mixture separated by a flat interface is quiescent and isotropic, with $\mathbf{v}=0$ and $Q_{ij}=0$. To lowest order in spatial gradients, the linear dynamics of fluctuations in the alignment tensor is then simply governed  by
\eq
\partial_t Q_{ij}(\rbf) = -\frac{1}{\tau}Q_{ij}(\rbf) \mcm{+D\nabla^2 Q_{ij}}+ \lambda A_{ij}(\rbf) \;,
\label{eq:QIsotrpoicResponseBulk}
\eqq
where 
$\tau =\gamma/r$ is the nematic relaxation time. Our goal is again to project the dynamics of the alignment tensor onto the active/passive interface. \mc{This is, however, more challenging than in the case where the active liquid crystal is nematic ($r<0$) discussed in the previous section because, given the vanishing of the order parameter in the base state, the linear dynamics of its fluctuations does not couple to vorticity, but only to the strain rate. The latter vanishes at the interface. Thus we cannot neglect all $y$ dependence and equate the strain rate with its value at the interface. We therefore need a different} ansatz for the
 \mcm{dependence} of the nematic fluctuations \mcm{on the distance from the interface} in the active phase. 
\mcm{In the absence of activity, }
the flow generated by a fluctuation \mcm{$h(k,t)$ of the interface} is  \mcm{due entirely to capillary forces and is immediately obtained from Eq.}~\eqref{eq:vphi},  \mcm{with a resulting} strain rate $A^\phi_{ij}$ in the active component ($y>0$), given by 
\begin{align}
A^\phi_{xx}(k,y) &={-\frac{\sigma k^2}{4\eta}} |k|y e^{-\abs{k}y} h(k,t) \;,\\
A^\phi_{xy}(k,y) &=  {-\frac{\sigma k^2}{4\eta}} i ky e^{-\abs{k}y} h(k,t)\;.
\label{eq:strain_phi}
\end{align}
\mcm{In other words flows induced by capillary forces associated with interface fluctuations of wavenumber $k$ extend through an interfacial layer of thickness $|k|^{-1}$.} It is then clear from Eq.~\eqref{eq:QIsotrpoicResponseBulk} that, neglecting elasticity, the $y$ dependence of nematic tensor fluctuations generated by capillary forces will have the same form as that of the strain rate $A_{ij}^\phi$. This suggests the ansatz
\begin{align}
    Q_{xx}(k,y,t) &=  \abs{k}ye^{-\abs{k}y} q_{xx}(k,t) \;,
    \label{eq:Qxx_ansatz_m}\\
    Q_{xy}(k,y,t) &= i k ye^{-\abs{k}y} q_{xy}(k,t)\;, 
    \label{eq:Qxy_ansatz_m}
\end{align}
with $q_{ij}(k,t)$ to be determined as a self-consistent solution of Eq.~\eqref{eq:QIsotrpoicResponseBulk}.
We insert Eqs.~\eqref{eq:Qxx_ansatz_m} and \eqref{eq:Qxy_ansatz_m} into Eq.~\eqref{eq:QIsotrpoicResponseBulk} and into the active contribution to the flow velocity, Eq. \eqref{eq:vQ}. With this ansatz, the process to obtain the dispersion relation is as in the previous section, with two caveats. In this case, we cannot sinply decouple amplitude and angle fluctuations, since the amplitude is also small, and thus we have three degrees of freedom, $h$, $q_{xx}$ and $q_{xy}$. Additionally, we observe there is a mode that decouples from the dynamics of $h$, which is a linear combination of $q_{xx}$ and $q_{xy}$. Thus we arrive to the linear equations for $h$ and $\psi_{\pm} = (q_{xx}\pm q_{xy})/2$ (see derivation details in Appendix \ref{app_sec:isotropic})
\begin{align}
    \partial_t  h(k,t) &= -\dfrac{\sigma k^2 }{4\eta\abs{k}}  h(k,t) -  \dfrac{\alpha}{4\eta k} \psi_+(k,t)\;, \label{eq:h_Isotropic_Linear}\\
    \partial_t \psi_+(k,t) &= - \left( \dfrac{1}{\tau}+\dfrac{\alpha \lambda}{4\eta}  + 2D k^2 \right)\psi_+(k,t)
\notag \\
    &- \dfrac{\sigma \lambda k^2}{4\eta}  h(k,t)  + \dfrac{\alpha \lambda}{4\eta}\psi_-(k,t)\;,\label{eq:Psi+_Isotropic_Linear}\\
     \partial_t \psi_-(k,t) &= -\left(\dfrac{\alpha \lambda}{4\eta} +\dfrac{1}{\tau}+ 2Dk^2 \right)\psi_-(k,t)\;.
\label{eq:Psi-_Isotropic_Linear}
\end{align}
\begin{figure}
    \includegraphics[width=0.49\textwidth]{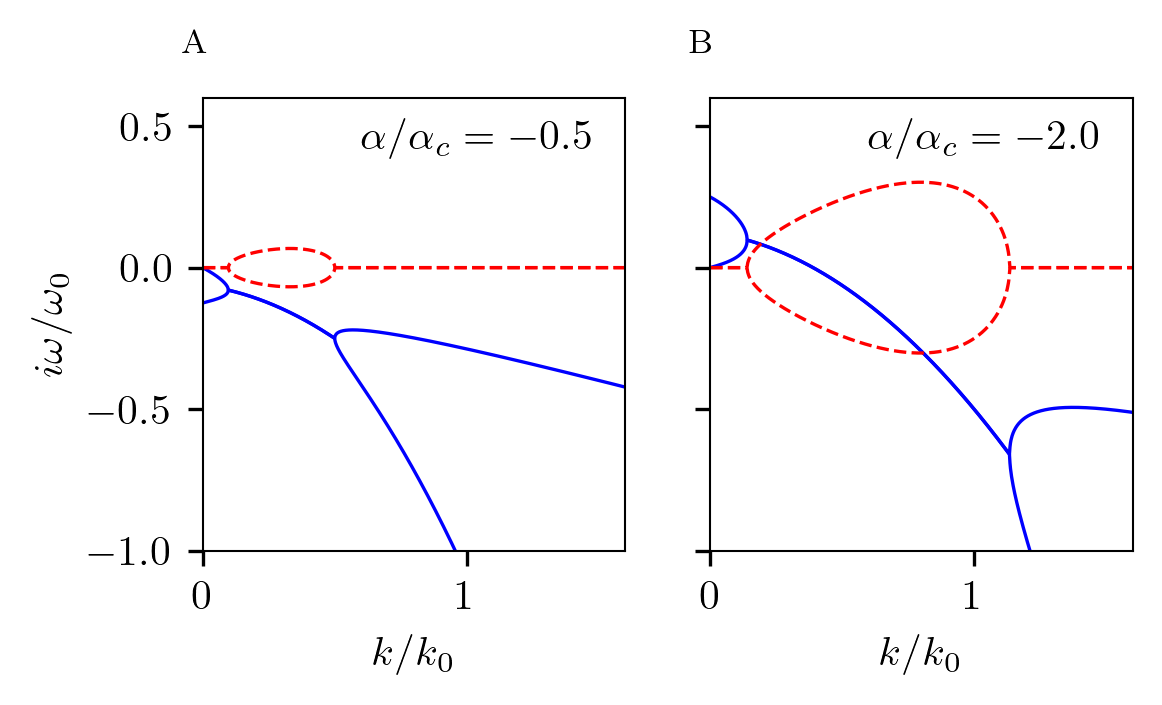}
    \caption{\mc{Dispersion relation for the modes controlling the dynamics of fluctuations of an active/passive interface with $r>0$. 
    Activity here is scaled by the critical threshold $\alpha_c$ 
    for the instability of the homogeneous state.  
    (A) For $|\alpha|<\alpha_c$ the modes are stable at all wavenumbers and propagating in a range of $k$. (B) For $|\alpha|>\alpha_c$ the modes are unstable at long wavelengths. Parameter values: $\lambda=1.0, \tau=4.0/\omega_0$,} which corresponds to \pg{$\alpha_c=\alpha_0$}.  }
\label{fig:isotropicdispersion}
\end{figure}

The dispersion relations corresponding to the previous equations are shown in Fig.~\ref{fig:isotropicdispersion}. As before, in absence of activity, all fluctuations decay. Again as before, extensile activity renders the isotropic state with flat interface unstable  when the rate $\sim |\alpha|/\eta$ at which active energy is injected exceeds the rate $1/\tau$ at which it is dissipated by rotational viscosity. The instability occurs at a critical activity $\alpha_c=4\eta/\lambda\tau$ in a band of wavevectors $k\in[0,k_c]$, with $k_c=\sqrt{\lambda(|\alpha|-\alpha_c)/4\eta D}$. For values of activity below $\alpha_c$, the interface can support \mc{stable} propagating modes (see Fig.~\ref{fig:isotropicdispersion}A). 
Wave propagation sets in for $k\geq k_{l}$, with
\begin{align}
    k_l &= \frac{4\eta}{\sigma \tau}\left(1 - \sqrt{\abs{\alpha}/\alpha_c}\right)^2\;.
\end{align}
As $\alpha$ approaches $\alpha_c$,  $k_l \rightarrow 0$ and we get a large scale instability. As activity keeps increasing, $k_l$ again becomes finite, and fluctuations larger than $1/k_l$ do not propagate along the interface. Close to the critical point $k_l \sim (1-|\alpha|/\alpha_c)^2.$

Qualitatively, therefore, whether the liquid crystal is in its isotropic or nematic regime does not change the linear phenomenology of the interface, which displays interface instabilities, connected to the bending instability of the liquid crystal, and wave propagation. The difference in the isotropic regime is \pgf{that the liquid crystalline degrees of freedom are not hydrodynamic, i.e, $\omega (k\rightarrow 0)\ne 0$. There is instead a relaxational rate} 
$1/\tau$ in Eq.~\eqref{eq:Psi+_Isotropic_Linear}, which causes the instabilities to now appear above a finite activity $\alpha_c$.

As in the nematic regime described in the previous section, the emergence of propagating waves in this system comes from the interplay between an active flow and interface tension. Activity generates spontaneous flows that create local nematic order, which destabilises the interface. The flow created by relaxation due to interface tension in turn drives the liquid crystal. Again, this nonreciprocity points to propagating structures and traveling waves in the system \cite{brauns2023non}.

\subsection{Effect of gravity\label{sec:gravity}}

\begin{figure}[t]
    \centering    \includegraphics[width=0.49\textwidth]{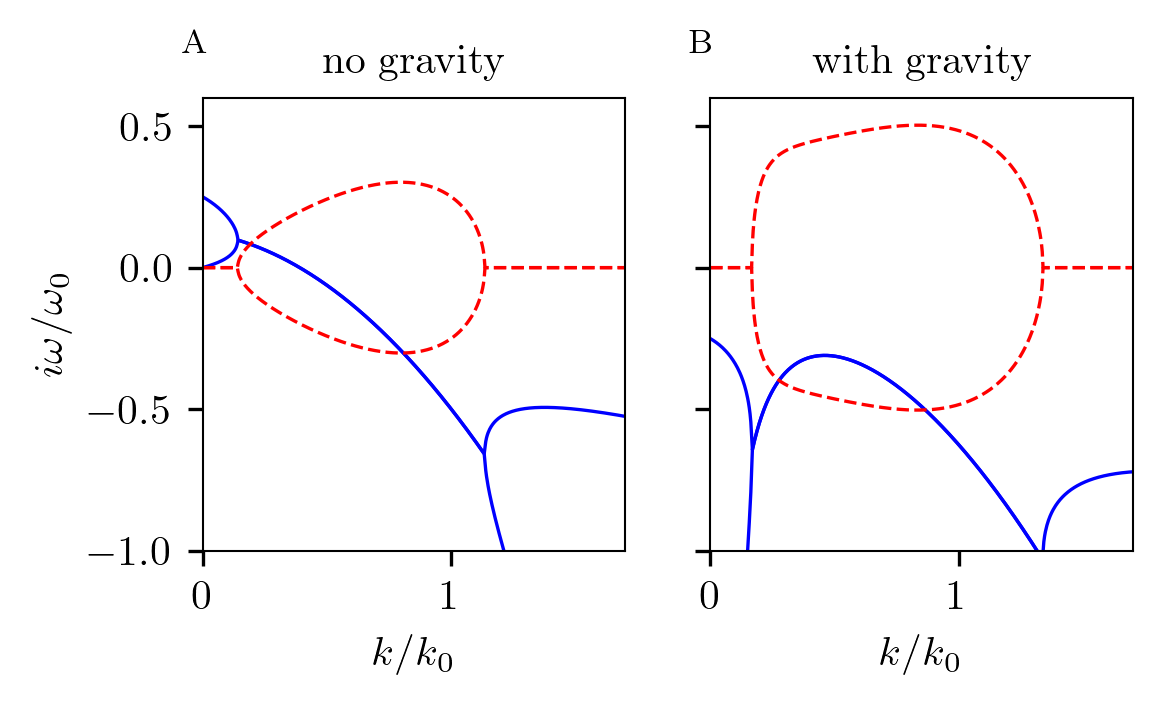}
    \caption{\mc{Dispersion relation for the dynamics of fluctuations of an active isotropic liquid crystal ($r>0$) and a passive fluid,  (A) without and (B) with gravity ($\ell_c k_0 =1.0$). All other parameters are as in Fig.~\ref{fig:isotropicdispersion}B, which is repeated here for ease of comparison. Gravity stabilizes the interfacial instability, while still allowing for propagating waves.} }
    \label{fig:disp_gravity}
\end{figure}

As we have noted in the previous sections, the bend instability of the bulk liquid crystal \mc{destabilizes and} \mc{eventually breaks} up the interface, as  observed in both experiments~\cite{tayar2023controlling} and simulations~\cite{caballero2022activity}. \mc{Experiments} have studied interface fluctuations \mc{in the presence of} 
gravity which provides a restoring force that \mc{prevents interface break up, allowing direct measurements of interfacial fluctuations.}

The stabilising effect of gravity can be added to our \mc{model as a body force proportional to the density of the fluid in the Stokes equation, Eq.~\eqref{eq:flow_generic}, given by}
\begin{equation}
    \mathbf{f}_g = \left( \rho_p(1-\phi) + \rho_a \phi  \right) g \mathbf{\hat{y}}\;,
\end{equation}
where $\rho_p$ and $\rho_a$ are the densities of the bulk (passive and active) phases. \mc{In current experimental realizations~\cite{adkins2022dynamics},} the active phase is denser, and settles to the bottom.  

\mc{The addition of gravity simply shifts the contribution from interfacial tension according to
\begin{align}
   \sigma k^2\rightarrow\sigma k^2 + g\Delta\rho\equiv\sigma(k^2+\ell_c^{-2})\;, 
\end{align}
where $\Delta \rho = \rho_a-\rho_p$.  The capillary length $\ell_c = \sqrt{\sigma/(g \Delta \rho)}$ cuts off the  instability of the interface at long wavelengths, as shown in Fig.~\ref{fig:disp_gravity}.}

\section{Numerical results}
\label{sec:numerics}
To test our linear theory, we have evaluated numerically the dynamical structure factor of \mc{the interface using} 
the full hydrodynamic theory, i.e. Eqs.~(\ref{eq:phi_full}-\ref{eq:StokesFlow}). The structure factor is given by
\begin{align}   S(k,\omega)=\int_{0}^\infty dt ~e^{i\omega t}\langle h(k,t+t_0)h(-k,t_0)\rangle\;,\label{eq:struct_factor}
\end{align}
where the brackets denote an average over different time windows, as described below.

\begin{figure}[b]
\includegraphics[width=0.48\textwidth]{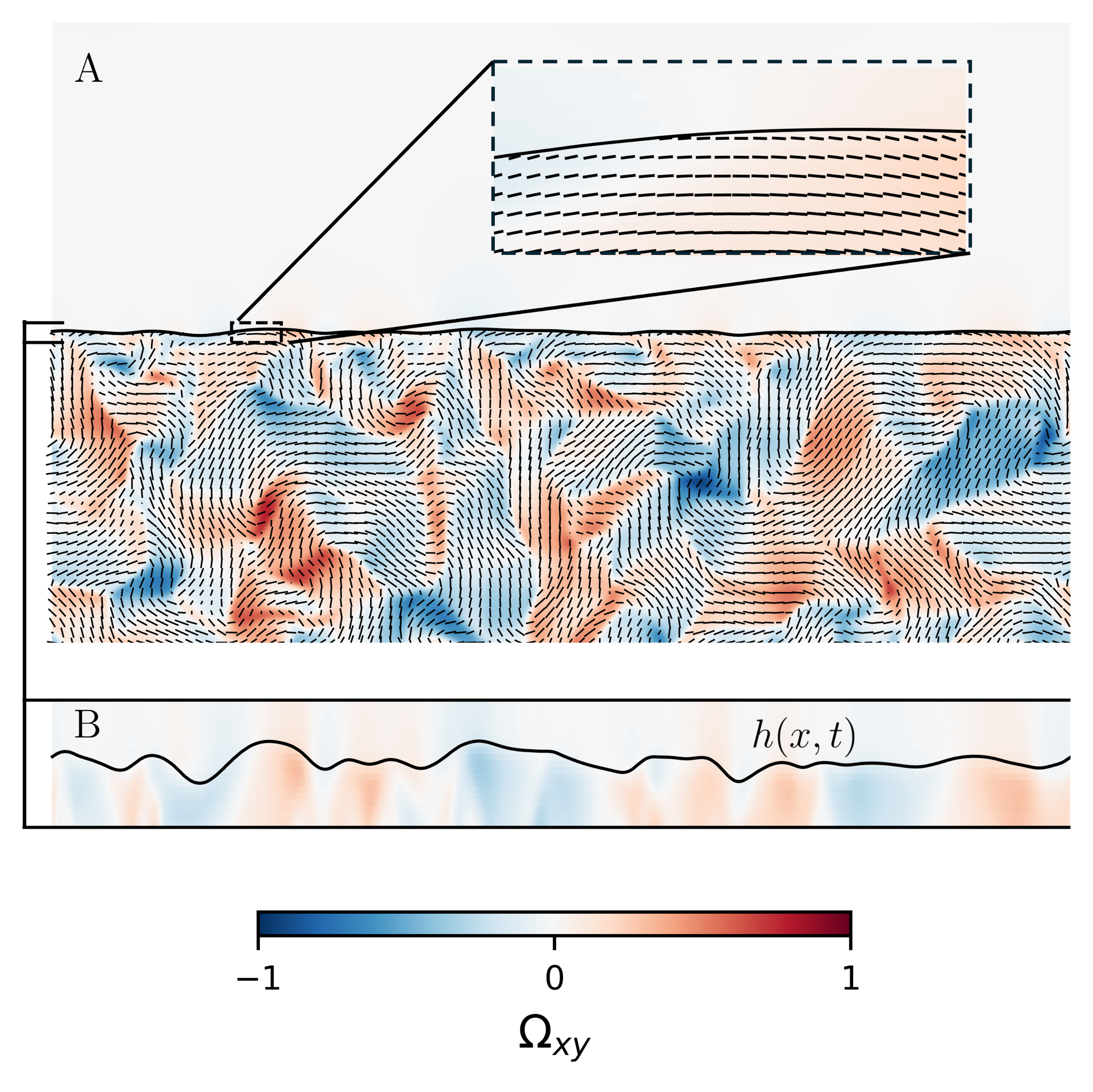}
\caption{ \revise{(A)}
 Snapshot of the phase separated system obtained by numerical integration of Eqs.~(\ref{eq:phi_full}-\ref{eq:StokesFlow}), \revise{with an isotropic liquid crystal ($r>0$)}. The director is plotted every \revise{ten} lattice sites for clarity \revfinal{(the length corresponds to the magnitude of the nematic order)}. \revfinal{Unbound topological defects are evident in the bulk, as expected in the regime of active turbulence.} \revise{The background color represents the \revfinal{(normalized) vorticity $\Omega_{xy}.$}} \revise{The inset shows a zoomed in region, highlighting the representative director structure (now shown every two lattice sites) close to the interface
 (black line). Simulation parameters: $K=10.0, a=-50, M=0.67, u=10.0, \alpha=-10.0, \lambda =1.0, \eta=1.0$ and $g\Delta \rho  = 1.2.$ (B) Zoomed in region showing the interfacial fluctuations quantified by the height field $h(x,t)$.}
}     
\label{fig:interface_simulations}
\end{figure}

\begin{figure*}
\includegraphics[width=\textwidth]{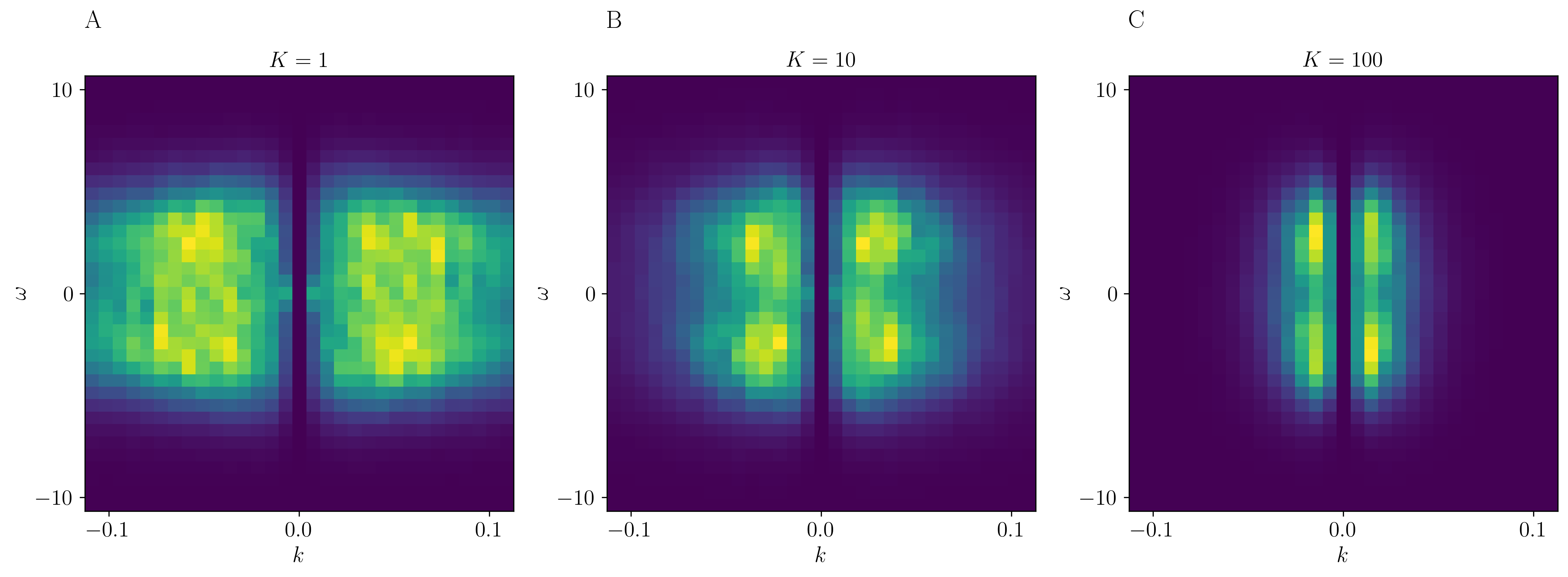}
\caption{\mc{The dynamic structure factor $S(k, \omega)$ of the interface computed numerically shows clear evidence of} propagating waves 
\mc{at intermediate wavenumbers. The three frames correspond to three values of} the nematic elastic \mc{stiffness, $K=1, 10, 100$. Increasing $K$ suppresses wave propagation at large} wavenumbers. The other parameters here are fixed to be: 
$a=-50, M=0.67, u=10.0, \alpha=-10.0, \lambda =1.0, \eta=1.0$ and $g\Delta \rho  = 1.2.$}     
\label{fig:simulation_strcture_factor}
\end{figure*}

To compute Eq.~\eqref{eq:struct_factor} we have used a pseudospectral code to numerically integrate Eqs.~(\ref{eq:phi_full}-\ref{eq:StokesFlow}) on a discrete lattice of size $N\times N$ for $N=1024$. The pseudospectral method provides a stable technique for dealing with high order gradients, as present in the Cahn-Hilliard equation. 
The integration has been \mc{carried out} with available pseudospectral solvers~\cite{caballero2024cupss}, using periodic boundary conditions in the $x$ direction. We enforce Dirichlet boundary conditions at the top and bottom boundaries by setting a fixed buffer zone in a band of width $128$ lattice sites at the top of the system in which  $\phi=0$ and $\mathbf{v}=0$, and a similar buffer zone at the bottom of the system where $\phi=1$ and $\mathbf{v}=0$. We report all the simulation parameters in physical units: lengths in units of the interface width $\ell_\phi,$ time in units of the nematic reorientation time $\gamma/r$ and stresses in units of the nematic condensation energy $r$.

The initial state is chosen as fully phase separated with a horizontal interface \mc{at $y=0$,} corresponding to $\phi(\textbf{r},t=0) = \left[1+\tanh \left(y/\sqrt{2}\ell_\phi\right)\right]/2$, so that the active liquid crystal is in the region $ y>0,$ and the passive fluid in the region $y<0$. \fer{The components of the order parameter $Q_{ij}$ are initialized as small noisy fluctuations around their mean value of zero, which allows activity to induce alignment.} 

The chosen values of the parameters are such that the active bulk develops a turbulent regime, \mc{but low enough for the interface to remain a single-valued function.} \mc{We additionally include a gravitational force to ensure a stable interface.} Active flows in the bulk  then drive the dynamics of the interface, which will fluctuate as a result of the interplay between interfacial tension and the destabilising effects of activity. The chosen value of activity is larger than the critical activity above which the isotropic state becomes unstable, but kept low enough so that interfacial tension and gravity keep a well-defined interface \revise{(Fig.~\ref{fig:interface_simulations}A)}.

In each integration of the model, we extract the interface by fitting each constant-$x$ slice of the field $\phi(x,y,t)$ to a function of the form 
\pg{$f(y) =  a \left( 1+\tanh [b(y-y_0)/ \sqrt{2} \ell_\phi] \right)$}
and use the resulting fitting parameter $h(x, t) = y_0$ as the location of the interface at point $x$ and time $t$ \revise{(Fig.~\ref{fig:interface_simulations}B)}. 
The dynamical properties of the interface are then extracted from $h(x,t)$ by calculating its structure factor $S(x,t) = \langle h(x,t) h(0,0) \rangle$, which we analyse in Fourier space $S(k,\omega) = | h(k,\omega)|^2$. 
We run the simulation for $2\cdot 10^7$ time steps, with $\Delta t = 10^{-4} \gamma/r$, i.e., total time $T =  2\cdot 10^3 \gamma/r$. This total time is much longer than the observed time scale of wave propagation. To extract a structure factor, we split this dataset into nonoverlaping time series, each containing the interface profile $h_i(x, t)$ with $t\in [t_i, t_i + \Delta T)$, with $\Delta T=10 \gamma/r$. This produces 200 time series. We then average the structure factor calculated from each of these intervals. 

\mc{The structure factor is shown in Fig. \ref{fig:simulation_strcture_factor} for three values of the nematic stiffness $K$}. \mc{Even though flows in the active fluid are chaotic and strongly nonlinear, the dispersion relation provides a qualitative good fit} for the observed frequencies and wavelengths \mc{of the capillary waves.}

\section{Discussion}

\mc{Activity has profound effects on phase separation, driving giant interfacial fluctuations, arresting coarsening, and perhaps providing a new handle for controlling the properties of interfaces~\cite{adkins2022dynamics,tayar2023controlling}. Many realizations of active fluids are composed of elongated entities with liquid crystalline degrees of freedom that play a key role in driving active flows. It is therefore important to understand the interplay of activity and orientational order on soft fluid-fluid interfaces. In these systems} interfacial fluctuations are driven by the chaotic dynamics of the bulk,  offering an opportunity to probe the properties of active liquid crystal by probing the behavior of the interface. 

In this paper, starting with a familiar continuum model of active liquid crystal hydrodynamics, we have derived analytically 
 linear equations for the behavior of fluctuations of the active/passive interface. \revise{This study builds on classical frameworks of interfacial waves in equilibrium ~\cite{langevin1992light, safran2018statistical}, which we extend to analyze the effects of active driving.} The derivation becomes \mc{challenging because} interfacial fluctuations are driven by active \mc{forces throughout the bulk fluid and} propagated nonlocally through the hydrodynamic flows. \mc{This is to be contrasted to scalar models of active phase separation, where active stresses only act at the interface~\cite{tiribocchi2015active}. To make progress, we have proposed}  simple approximations that qualitatively capture much of the phenomenology seen in experiments \mc{and in numerical simulations of the full nonlinear hydrodynamic model.}

We believe our work will help understand the effect of activity on interfacial dynamics and offer tools for quantitative comparison with experiments. \revise{
Traveling waves at the surface of a phase separated active microtubule-kinesin suspension were observed in Ref.~\cite{adkins2022dynamics} using a quasi-2D system with very low surface tension. It was also shown there that, although those experiments are strictly outside the range of application of a linear theory even at the lowest kinesin concentrations, a simplified version of the present theory does reproduce the qualitative features of the dispersion relations. On the other hand, a theory of the type presented here cannot provide a quantitative prediction of the amplitude of fluctuations, which  has to be treated as a fitting parameter.  
Very recent experiments on active phase separation in bulk fluids with a 2D interface have also revealed traveling surface waves~\cite{zhao2024asymmetric} and may offer the opportunity for a more quantitative comparison of the structure of the dispersion relations. This, together with an extension to the nonlinear regime, provides an important direction}
for future work.

A clear direction is expanding this theory is to incorporate nonlinear terms. This has been done in the past in the context of nonequilibrium dynamics~\cite{kardar1986dynamic}, and more recently as applied to scalar active systems \cite{caballero2018strong, besse2023interface}, \mc{where active forces only act directly at the interface. Extending this work to situations where} active hydrodynamic flows throughout the bulk are the main drive of interfacial dynamics \mc{remains an open challenge. More work is needed to establish whether} active hydrodynamic interfaces belong to the same universality class as other interface models used in the past. 

Another future direction  is expanding this model to more complete theories that consider two  flows coupled by friction, as has been done before in related studies of binary mixtures \cite{bhattacharyya2023phase, tanaka2000viscoelastic, weber2018differential}. It is unknown, however, if this more complete description would provide a deeper insight into the complicated phenomenological aspect of the experimental systems that have inspired this work \cite{adkins2022dynamics}.

\mc{While these open questions are challenging, it is clear that a quantitative understanding of the dynamics of complex active interfaces will be required for the development of new functional soft materials where the interfacial properties are tuned and controlled by active processes.}

\section*{Conflicts of Interest}
There are no conflicts of interest to declare.

\section*{Acknowledgements}
We would like to thank Cesare Nardini, Liang Zhao and Austin Hopkins for illuminating conversations.

The work on the isotropic liquid crystal/passive fluid interface (PG, IK, ZY) and the numerical work (PG) were supported by the U. S. Department of Energy Office of Basic Energy Sciences (DE-SC0019733). The work on nematic/passive interface (FC, MCM) was supported by National Science Foundation (DMR-2041459). All authors contributed to the writing of the paper. Use was made of computational facilities purchased with funds from the National Science Foundation (CNS-1725797) and administered by the Center for Scientific Computing (CSC). The CSC is supported by the California NanoSystems Institute and the Materials Research Science and Engineering Center (MRSEC; NSF DMR 2308708) at UC Santa Barbara.

\appendix
\begin{widetext}
\section{Interface Projection}
\label{app_sec:interfaceprojection}
This appendix shows the  calculation of the  interface dynamics following the method of Ref.~\cite{bray2001interface}.
We start with the (incompressible) Stokes flow generated by the force $\mathbf{F}$, which is given by
\begin{align}\label{eq:eq1}
    v_i(\mathbf{r}) &=\int d\mathbf{r'}\, T_{ij}(\mathbf{r-r'})F_j(\mathbf{r'})\; ,
\end{align}
where $T_{ij}$ is the Oseen tensor \mcm{given by Eq.~\eqref{eq:Tq} and}
\begin{equation}
\mcm{\mathbf{F}} = \mu \nabla \phi +\alpha\nabla\cdot ({\phi}\mathbf{Q}) \; ,    
\end{equation}
where the chemical potential can be written in terms of the free energy density of the phase field, $\mu = f_0'(\phi) - \kappa \nabla^2\phi$.
Here, we write the capillary force for the bulk system to be equal to $\mu\nabla \phi$,  integrating by-parts $\nabla\cdot \sigma^\phi = -\phi \nabla \mu$ and ignoring the boundary terms with vanishing contribution to the flow.

We consider the initial phase separated state and height fluctuations $h\mcm{(x,t)}$ of the flat interface. \mcm{We can write} $\phi(x,y,t) =  g(u)$, where $u=y - h(x,t)$ is the distance from the interface and $g(u)$ \mcm{varies sharply at the interface}. We then obtain

\begin{align}
\boldsymbol{\nabla} \phi &= g'(u)\left(\hat{\mathbf{e}}_y 
 -\hat{\mathbf{e}}_x (\partial_x h) \right), \\
 \nabla^2\phi &= \left(1 + (\partial_x h)^2\right)g''(u) - g'(u)\partial_x^2h \; ,
\end{align}
where the prime denotes a derivative. The capillary force is \mcm{then given by}
\begin{equation}\label{eq:mugradphi}
\begin{aligned}
    \mu \boldsymbol{\nabla} \phi  &=  g'(u)\left( \hat{\mathbf{e}}_y 
    -\hat{\mathbf{e}}_x (\partial_x h)
    \right) \bigg(f_0'(g) -\kappa \big( (1+(\partial_x h)^2)g''(u) -g'(u) \partial_x^2h \big)\bigg) \\
    &= \left(\hat{\mathbf{e}}_y 
    -\hat{\mathbf{e}}_x (\partial_x h)
    \right) \bigg( \frac{d f_0}{d u} -\frac{\kappa}{2} \left(1+(\partial_x h)^2 \right ) \frac{d [g'(u)]^2}{d u } +\kappa [g'(u)]^2 \partial_x^2h \bigg)\;.
\end{aligned}
\end{equation}
The passive flow $v_i^p(x,y)$ can be calculated from Eqs.~\eqref{eq:eq1} and \eqref{eq:mugradphi}, with the result (to linear order in fluctuations)
\begin{align}
    v_i^\phi(x,y) = {\sigma} \int_{-\infty}^\infty dx' \; T_{iy}(x-x', y) \partial_{x'}^2 h(x',t)\;,
\end{align}
where $\sigma = \kappa \int du\, [g'(u)]^2$ is the surface tension. The terms proportional to $f_0'(u)$ and $d(g')^2/du$ vanish upon integrating by parts because they are total derivatives.

To evaluate the flows explicitly it is useful to introduce the partial Fourier transform of the Oseen tensor as 
\begin{align}
    G_{ij}(k,y) = \frac{1}{2\pi}\int_{-\infty}^{\infty}\, dk_y e^{i k_y y} T_{ij}(k, k_y)\;.
\end{align}
These partial Fourier transforms can be easily evaluated:
\begin{align}
G_{xx}(k, y) &= \frac{e^{-|ky|}}{4\eta |k|}\left(1- |ky|\right)\;,\notag\\
G_{xy}(k, y) &= -i ky \frac{e^{-|ky|}}{4\eta |k|}\;,\notag\\
G_{yy}(k, y) &= \frac{e^{-|ky|}}{4\eta |k|}\left(1+ |ky|\right)\;. \label{eq_app:G_zero_friction}
\end{align}
Flows of wavenumber $k$ decay on length scales $k^{-1}$ and are only cutoff by the system size.

The  \textit{passive flows} due to the surface tension are then given by
\begin{align}
    v_i^\phi(k, y) &=  -\sigma k^2 G_{iy}(k,y) h(k, t)\;. \label{eq:passivelinearflows}
\end{align}
or explicitly,
\begin{align}
    v_x^\phi(k, y)&= \frac{\sigma|k|}{4\eta}e^{-|ky|}(iky) h(k, t) \;,\\
    v_y^\phi(k, y)&= -\frac{\sigma |k|}{4\eta}(1+|ky|)e^{-|ky|} h(k, t)\;.
\end{align}

The active flows can be written as
\begin{align}
v_i^a(x,y) = \alpha \int_{-\infty}^\infty dx'  \int_{-\infty}^\infty dy'\;& T_{ij}(x-x', y-y') \; 
\partial_{k}' \left( \phi(x',y') Q_{jk}(x', y')\right)\;,   
\label{eq:va-gen}
\end{align}
Here, $\phi(x,y,t') = g(y - h(x, t)),$ where $g$ varies sharply from $0$ to $1$ at the interface. To linear order in fluctuations, $Q_{ij} = Q^0_{ij} + \delta Q_{ij}$, where the explicit form of the  base state $Q^0_{ij}$ depends on whether the active liquid crystal is in the isotropic or nematic state when passive.  

It is easy to show that the active flow always vanishes in the base state, i.e., to zeroth order in fluctuations. This is obvious in the isotropic  state where $Q_{ij}^0=0$.  In the nematic case the base state has components
$Q^0_{xx}=-Q^0_{yy}=S_0(y)$, where $S_0(y)$ describes the decay of nematic order across the interface, and $Q^0_{xy}=0$, which corresponds to a pressure jump across the interface, and therefore produce no flow.

Transforming Eq.~\eqref{eq:va-gen} to Fourier space in $x$, and expanding to linear order in fluctuations, the active flow can then be written as
\begin{equation}
    \begin{aligned}
        &v^a_i(k,y) = -\alpha ik h(k)\int_{-\infty}^\infty dy' G_{ij}(k,y-y'){g}'(y')Q_{jx}^0(y') + \alpha\int_{-\infty}^\infty dy' G_{ij}(k,y-y'){g}'(y') \mcm{\delta}Q_{jy}(k,y') \\
        & + \alpha\int_{-\infty}^\infty dy' G_{ij}(k,y-y'){g}(y')\left[ik\delta Q_{jx}(k,y')+\partial'_y\delta Q_{jy}(k,y')\right] - \alpha\int_{-\infty}^\infty dy' G_{iy}(k,y-y'){g}''(y')h(k)Q_{yy}^0(y').
    \end{aligned}
\label{eq_app:v_active_general}
\end{equation}
In the limit when the interface width is small, we can approximate the step function ${g}(y) = \Theta(y)$ and ${g}'(y) = \delta(y)$.  
Then we can simplify the active flow to be: 
\begin{equation}
    \begin{aligned}
        &v_i^a(k,y) =   -\alpha h(k)\left[ik G_{ix}(k, y)\mcm{-}\partial_yG_{iy}(k,y)\right]S_0(0) \\              &+\alpha\int_0^{\infty} dy' \left[ ikG_{iy}(k,y-y')+\partial_yG_{ix}(k,y-y')\right]\delta Q_{xy}(k, y')
        \\
        &+\alpha\int_0^{\infty} dy' \left[ ikG_{ix}(k,y-y')-\partial_yG_{iy}(k,y-y')\right]\delta Q_{xx}(k, y')\;.
     \label{eq:activeflows_general}
    \end{aligned}
\end{equation}

\section{Ordered Nematic}
\label{app_sec:ordered_nematic}
This section shows details on the derivation of Eqs.~\eqref{eq:hthetaordering_1} and~\eqref{eq:hthetaordering_2}, and the corresponding dispersion relation.

As described in the previous section, the base state here has $Q_{xx}^0 = -Q_{yy}^0 = S_0(y)$ and $Q_{xy}^0 = Q_{yx}^0 =0,$ where $S(y)$ describes the decay of nematic order across the interface, and thus correponds to a pressure jump. We further consider this in the limit with small nematic correlation length and treat the base state as a step function with a finite magnitude of the order parameter at the interface.

On top of this base state we consider fluctuations in the angle of the nematic director, i.e:
\begin{align}
    \delta Q_{xy}(k,y) = & S_0(y) \theta(k, y, t), \notag\\ 
    = &\theta(k,y,t) \quad \text{for } y>0
\end{align}
and then we can write the resulting active flows to linear order in fluctuations using Eq.~\eqref{eq:activeflows_general}. Using the aligned nematic as the base state and consider the angle fluctuations, we have:
 \begin{align}
    &v_i^a(k,y) = -\alpha h(k,t) \left[ik G_{ix}(k, y) \mcm{-} \partial_y G_{iy}(k,y)\right] S_0(0)\notag \\ &+ \alpha\int_0^{\infty} dy'\;\left[ ik G_{iy}(k, y-y')+\partial_y G_{ix}(k, y-y') \right] \theta(k, y', t).
\label{eq_app:v_active_theta}
 \end{align}

As we discuss in the main text, to get explicit dispersion relation coupling the interface and nematic fluctuations, we assume that the angular fluctuations are correlated over a length $\ell$ into the active phase i.e. $\theta(k, y',t) = \theta(k, t) e^{-y'/\ell}$. Using these we can explicitly calculate the active flows, \mc{with the result}
\pg{
\begin{align}
 v_x^a(k, y)  & = - \frac{\alpha S_0}{2\eta |k|}e^{-|ky|}\left(1 - |ky| \right)i k h +\frac{\alpha \ell}{\eta}\left[ e^{-y/\ell} \frac{k^2\ell^2+1}{(k^2\ell^2-1)^2} - e^{-|ky|}\frac{(1+|k|\ell)^2(1- |k|y + k^2 \ell y)}{2(k^2\ell^2-1)^2}\right]\theta\;, \\
 v_y^a(k, y) &= -\frac{\alpha S_0}{2\eta}e^{-|k|y}(|k| y) h + \frac{\alpha \ell^2}{\eta} \left[e^{-y/\ell} \frac{k^2\ell^2+1}{(k^2\ell^2-1)^2} - e^{-|k y|}\frac{(1+|k|\ell)^2(1-y/\ell + |k|y)}{2(k^2\ell^2-1)^2}\right] ik\theta\;.
\end{align}
}
From these active flows, we can then calculate the strain rate and the vorticity.

\mc{By combining contributions from active and passive flows, we can then obtain equations for the coupled dynamics of height and order parameter.} 
To obtain an equation for the director angle fluctuations, we linearize the dynamics of $Q_{xy} \sim \sin(2\theta)$ and then multiply with a sharply peaked function ${g}'(y)$, integrating out the $y$-dependence. In the limit of a sharp interface, when ${g}'(y)\sim \delta(y)$, we \mc{obtain}
\begin{align}
    \partial_t h(k, t) &= v_y |_{y\rightarrow0} \;,\\
    \partial_t \theta(k, t) &= \lambda A_{xy}|_{y\rightarrow0} + \Omega_{xy} |_{y\rightarrow0} - D \left(k^2+ \frac{1}{\ell^2}\right)\theta(k,t),
\end{align}
where the limit should be considered as the mean from both sides of the interface i.e. $x|_{y\rightarrow 0} = \left(x|_{y\rightarrow 0^+} + x|_{y\rightarrow 0^-}\right)/2.$

\mc{Inserting the expressions for active and} passive flows computed earlier, we finally obtain the following coupled equations
\pg{\begin{align}
    \partial_t h &= -\frac{\sigma}{4\eta}\abs{k} h + \frac{\alpha}{2\eta} \frac{\ell^2}{(1+|k|\ell)^2} ik\theta\;, \label{eq_app:orderednematic_ell_h}\\
    \partial_t \theta &= -\frac{\sigma \abs{k}}{4\eta}ikh - \frac{\alpha}{2\eta} \theta \left[ \frac{\lambda+1}{2}\frac{k^2\ell^2}{(1+|k|\ell)^2} + \frac{\lambda-1}{2}\left( 1 + \frac{1}{(1+|k|\ell)^2}\right) \right] - D \left(k^2+ \frac{1}{\ell^2}\right) \revise{\theta}\;.\label{eq_app:orderednematic_ell_theta}
\end{align}}
For a thin nematic layer, $k\ell \ll1$ these reduce to,
\begin{align}
    \partial_t h &= -\frac{\sigma}{4\eta}\abs{k} h + \frac{\alpha \ell^2}{2\eta} ik\theta \;,\\
    \partial_t \theta &= -\frac{\sigma \abs{k}}{4\eta}ikh - \left(\frac{\alpha (\lambda-1)}{4\eta} + \frac{D}{\ell^2} \right) \theta\;,
\end{align}
\mc{Conversely,} in the limit $\ell \rightarrow\infty$ \mc{we recover  Eqs.~\eqref{eq:hthetaordering_1} and \eqref{eq:hthetaordering_2} of the main text.}

\begin{figure*}
    \centering
    \includegraphics[width=\textwidth]{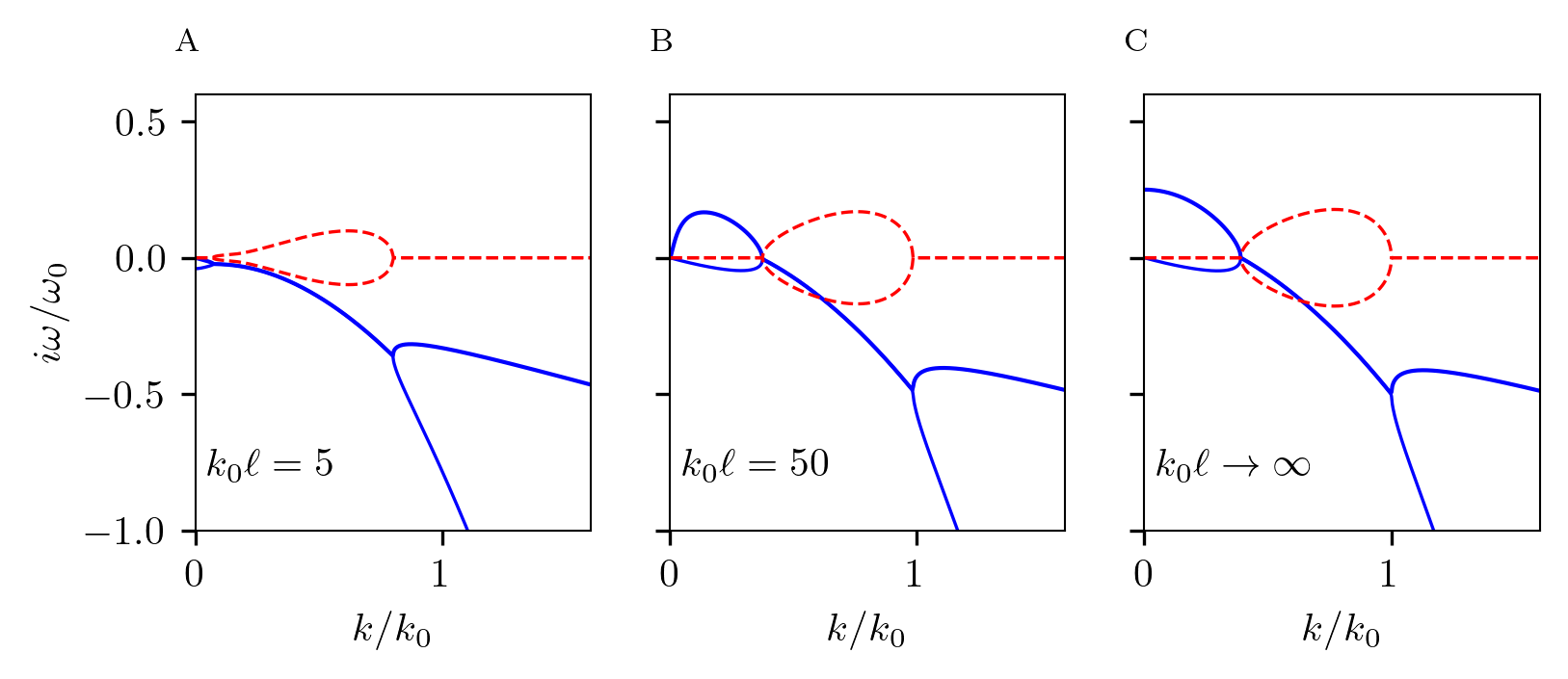}
    \caption{Dispersion \mc{relations obtained from Eqs.~\eqref{eq_app:orderednematic_ell_h} and \eqref{eq_app:orderednematic_ell_theta} for different values of the length $\ell$.}
  As in the main text,  frequency, wavenumber and activity \mc{are scaled with $\omega_0 = \sigma^2/(\eta^2 D)$, $k_0 = \sigma/(\eta D)$, and $ \alpha_0 = \sigma^2/(D\eta)$.}  \mc{(A,B) A finite correlation length $\ell$ suppresses the interfacial instability, while still allowing for propagating modes. (C) For $\ell \rightarrow \infty$, we recover the dispersion relations shown in the main text. Parameter values: $\lambda =1.0$ and $\alpha = -\alpha_0/2$.} }
    \label{fig:disp_appendix}
\end{figure*}
    
\section{Isotropic Liquid Crystal}
\label{app_sec:isotropic}
For the isotropic liquid crystal ($r >0$), the base state is ${Q}^0_{ij} = 0,$ and then, using Eq.~\eqref{eq:activeflows_general}, the flows are simply given by:
\begin{equation}
    \begin{aligned}
        v_i^a(k,y) = & \alpha\int_0^{\infty} dy' \left[ ikG_{iy}(k,y-y')+\partial_yG_{ix}(k,y-y')\right]\delta Q_{xy}(k, y')
        \\
        +&\alpha\int_0^{\infty} dy' \left[ ikG_{ix}(k,y-y')-\partial_yG_{iy}(k,y-y')\right]\delta Q_{xx}(k, y')\;.
    \label{eq_app:isotropic_active_flows}
    \end{aligned}
\end{equation}
Here, $\delta Q_{ij}(y')$ are the fluctuations in the isotropic state. As before, from Eq.~\eqref{eq:passivelinearflows} we have the passive flows,
\begin{align}
    v_i^\phi = -\sigma k^2 G_{iy}(k,y) h(k, t)\;
\end{align}
This leads to the following strain rate (in the active phase, $y>0$), 
\begin{align}
A^\phi_{xx}(k,t) &=-\frac{\sigma k^2}{4\eta} |k|y e^{-\abs{k}y} h(k,t) \;,\\
A^\phi_{xy}(k,t) &=  -\frac{\sigma k^2}{4\eta} i ky e^{-\abs{k}y} h(k,t)\;.
\label{eq:strain_phi}
\end{align}
This informs our ansatz for the fluctuations in the nematic order
\begin{align}
    \delta Q_{xx}(k,y,t) &=  \abs{k}ye^{-\abs{k}y} q_{xx}(k,t) \;,
    \label{eq:Qxy_ansatz}\\
    \delta Q_{xy}(k,y,t) &= i k ye^{-\abs{k}y} q_{xy}(k,t)\;. 
    \label{eq:Qxx_ansatz}
\end{align}
The resulting active flows, using Eq.~\eqref{eq_app:isotropic_active_flows} are given by (in the active phase, $y>0)$
\begin{align}
    v_x^{a}(y) &= \mc{\dfrac{\alpha}{24\eta \abs{k}}ik y ~e^{-\abs{k}y}}\left[\left(3 + 6\abs{k}y -2k^2y^2\right)q_{xx} +\left(3 -6\abs{k}y +2k^2y^2\right)q_{xy} \right]\;, \\
    v_y^{a}(y) &= \mc{\dfrac{\alpha  }{24\eta \abs{k}}~e^{-\abs{k}y}}\left[\left(-3- 3\abs{k}y +2k^2\abs{k}y^3\right)q_{xx} -\left(3  +3 \abs{k} y +2k^2 \abs{k}y^3\right)q_{xy} \right]\;. 
\end{align}
\revise{From the flow, we can calculate the strain rate which can then be integrated} over $y$ in the active phase, $\int dy\,A^a_{ij}(k,y)$, to write the dynamics of the nematic order fluctuations 
\begin{align}
    \partial_t h(k,t) &= { \left(v_y^p(k)|_{y\rightarrow0} + v_y^a(k)|_{y\rightarrow0}\right)}\;,\\
    \partial_t q_{xx}(k,t) &= -\left(\frac{1}{\tau}+{2Dk^2}\right)q_{xx}(k,t) + \lambda |k| \int_0^\infty dy \, \left(A^\phi_{xx}(k, y) + A^a_{xx}(k, y)\right)\;,\\
    \partial_t q_{xy}(k,t) &= -\left(\frac{1}{\tau} +{2Dk^2}\right)q_{xy}(k,t) - i \lambda k \int_0^\infty dy \, \left(A^\phi_{xy}(k, y) + A^a_{xy}(k, y)\right)\;.
\end{align}
Using the active and passive flows calculated above, we get:
\begin{align}
     \partial_t h &= -\frac{\sigma k^2 }{4\eta\abs{k} } h - \frac{\alpha }{8 \eta |k|}  q_{xy} - \frac{\alpha}{8 \eta |k|}q_{xx}\;,\\
     \partial_t q_{xx} &= -\left( \frac{1}{\tau}+\frac{\alpha \lambda}{8\eta}  +{2Dk^2} \right) q_{xx}-\frac{\alpha \lambda }{8\eta}q_{xy}
     - \frac{\sigma \lambda k^2 }{4\eta}h \;,\\
     \partial_t q_{xy} &=  - \left(\frac{1}{\tau}+\frac{3\alpha \lambda}{8\eta} +{2Dk^2}\right)q_{xy} + \frac{\alpha \lambda }{8\eta} q_{xx} - \frac{\sigma \lambda k^2 }{4\eta} h\;.
\end{align} 
Finally, a simple change of variables decouples one of the above equations. We define
$\psi_\pm = \left(q_{xx} \pm q_{xy} \right)/2$ and obtain Eqs.~\eqref{eq:h_Isotropic_Linear}, \eqref{eq:Psi+_Isotropic_Linear} and \eqref{eq:Psi-_Isotropic_Linear} of the main text
\begin{align}
    \partial_t  
    \begin{pmatrix}
        h \\ \psi_+ \\ \psi_- 
    \end{pmatrix} =
    \begin{pmatrix}
     -\dfrac{\sigma k}{4\eta} & -\dfrac{\alpha}{4\eta k} & 0\\[2ex]
     -\dfrac{\sigma \lambda k^2}{4\eta}& -\dfrac{\alpha \lambda}{4\eta} -\dfrac{1}{\tau} -{2Dk^2}&\dfrac{\alpha \lambda}{4\eta}\\[2ex]
     0 & 0 & -\dfrac{\alpha \lambda}{4\eta} -\dfrac{1}{\tau} -{2Dk^2}
    \end{pmatrix}
    \begin{pmatrix}
        h \\ \psi_+ \\ \psi_- 
    \end{pmatrix}
\end{align}

\end{widetext}
\bibliographystyle{rsc} 
\bibliography{ref.bib}
\end{document}